\documentclass{tlp}

\usepackage{fancyvrb}
\usepackage{url}
\usepackage{amsmath}
\usepackage{mathpartir}
\makeatletter
\def\arcr{\@arraycr}
\makeatother
\usepackage{amssymb}

\usepackage{prooftree} 

\usepackage{multirow}
\usepackage{multicol}

\usepackage{xspace} 

\usepackage{float}
\floatstyle{boxed}
\restylefloat{figure}

\usepackage{flushend}

\newcommand{\GAMMAECVD}{E_c \vdash}

\title{Efficient Algebraic Effect Handlers for Prolog}
\author[A. H. Saleh and  T. Schrijvers]{AMR HANY SALEH \and TOM SCHRIJVERS\\
KU Leuven, Belgium \\
Department of Computer Science \\
\email{\{ah.saleh,tom.schrijvers\}@cs.kuleuven.be}
}

\begin{document}

\maketitle

\begin{abstract}
Recent work has provided \emph{delimited control} for Prolog to
dynamically manipulate the program control-flow, and to implement a wide range of
control-flow and dataflow effects on top of. Unfortunately, delimited control
is a rather primitive language feature that is not easy to use. 

As a remedy, this work introduces \emph{algebraic effect handlers} for Prolog,
as a high-level and structured way of defining new side-effects in a modular
fashion.  We illustrate the expressive power of the feature and provide an
implementation by means of elaboration into the delimited control primitives.

The latter add a non-negligible performance overhead when used extensively. To
address this issue, we present an optimised compilation approach that combines
partial evaluation with dedicated rewrite rules. The rewrite rules are driven
by a lightweight effect inference that analyses what effect operations may be called by a goal.
We illustrate the effectiveness of this approach on a range of
benchmarks. \emph{This article is under consideration for acceptance in TPLP}.
\end{abstract}


\section{Introduction}

The work of Schrijvers et al.~\shortcite{DBLP:journals/tplp/SchrijversDDW13} has introduced \textit{delimited control}
constructs for Prolog. Delimited control is a very powerful means to
dynamically manipulate the control-flow of programs that was first explored in
the setting of functional programming~\cite{Felleisen:1988,abstracting_control}.
Schrijvers et al.~\shortcite{DBLP:journals/tplp/SchrijversDDW13} show its
usefulness in Prolog to concisely define implicit state, DCGs and coroutines.
More recently, Desouter et al.~\shortcite{DBLP:journals/tplp/DesouterDS15} have shown that delimited control
also concisely captures the control-flow manipulation of tabling.

Unfortunately, there are two prominent downsides to delimited control.
Firstly, it is a rather primitive feature that has been likened to the
imperative \texttt{goto}, which was labeled harmful for high-level programming by
Dijkstra~\shortcite{Dijkstra:1968:LEG:362929.362947}. Secondly, the overhead of delimited control for encoding
state-passing features is non-negligible. For example, the delimited control implementation
of DCGs is 10 times slower for a tight loop than the traditional implementation.

This paper addresses both issues. In order to provide a high-level structured
interface to delimited control, we adapt the \emph{algebraic effects and
handlers} approach to Prolog. Algebraic effects and handlers have been said to
relate to delimited control the way structured loops relate to \texttt{goto}.
While the structured approach restricts the expressive power, we still show a
range of useful applications. Moreover, in exchange for the restricted
expressiveness, we provide two benefits. Firstly, multiple handlers can be
combined effortlessly to deal with distinct effetcs, to deal with one effect 
in terms of another or to customize the behavior of an effect.
Secondly, we provide an automated program transformation that eliminates
much of the overhead of delimited control. Indeed, compared to the free form of
delimited control, the structured approach of effect handlers simplifies the
identification of program patterns that can be optimised.

Our specific contributions are as follows:
\begin{itemize}
\item We define the syntax of algebraic effects and handlers for Prolog, and
      provide semantics in terms of an elaboration
      to the delimited control primitives.

\item We illustrate the feature with a number of examples, including DCGs, implicit
      state and a writer effect.

\item We provide a program transformation to eliminate much of the overhead of 
      the elaboration-based implementation. This transformation is formulated 
      in terms of partial evaluation augmented with rewrite rules. These
      rewrite rules are driven by an effect analysis that characterises which
      effects may be generated by a goal.

\item We have implemented our program analysis and illustrated its
      effectiveness on a range of benchmarks.

\end{itemize}
All code of this paper is available at \url{http://github.com/ah-saleh/prologhandlers}.

\section{Algebraic Effect Handlers}

This section introduces our algebraic effect handlers for Prolog.

\subsection{Syntax and Informal Semantics}\label{sec:syntax}

We introduce two new syntactic constructs. The
\emph{effect operations} are Prolog predicate symbols
$\mathit{op}$\texttt{/}$n$ that are declared as such with the following syntax.

\begin{Verbatim}[commandchars=\\\{\},xleftmargin=5mm]
:- effect \textit{op}/\textit{n}.
\end{Verbatim}

For instance, we declare operation \texttt{c/1} to consume a token,
\texttt{get/1} and \texttt{put/1} to respectively retrieve and overwrite
an implicit state, and \texttt{out/1} to output a term.

The \emph{handler} is a new Prolog goal form that specifies how to interpret
effect operations. Its syntax is as follows:
\[
\begin{array}{c}
             \texttt{handle } G_0  \texttt{ with} \\
             op_{1}(\bar{X}) \rightarrow G_{1};\\
             \dots\\
             op_{n}(\bar{X}) \rightarrow G_{m}\\
             \lbrack \texttt{finally}(G_f) \rbrack \\
             \lbrack \texttt{for}(P_1 = T_1,\ldots,P_n = T_n) \rbrack 
\end{array}
\]
Effect handlers can be thought of as a generalisation of exception handlers,
where calling an effect operation corresponds to throwing an exception.  The
handler ``catches'' the operations that arise in $G_0$. Its operation
clauses $\mathit{op_i}(\bar{X}) \rightarrow G_i$ stipulate that an occurrence
of operation $\mathit{op}_i(\bar{X})$ is to be handled by the goal $G_i$. 

Before we explain the optional \texttt{finally} and \texttt{for} clauses,
consider a few ways in which the \texttt{out/1} operation can be handled in
\texttt{hw/0}.
\begin{Verbatim}[xleftmargin=5mm]
  hw :- out(hello), out(world).
\end{Verbatim}
In terms of the exception analogy, \texttt{hw/0} throws two \texttt{out/1}
exceptions. Our first handler intercepts the first \texttt{out/1} and does nothing.
\begin{Verbatim}[framerule=1mm,frame=leftline,xleftmargin=5mm]
  ?- handle hw with (out(X) -> true).
  true.
\end{Verbatim}
A more interesting handler prints the argument of \texttt{out/1}.
\begin{Verbatim}[framerule=1mm,frame=leftline,xleftmargin=5mm]
  ?- handle hw with (out(X) -> writeln(X)).
  hello
  true.
\end{Verbatim}
Note that only the first \texttt{out/1} is handled; this aborts the remainder
of \texttt{hw} and the second \texttt{out/1} is never reached. To handle all operations,
effect handlers support a feature akin to \emph{resumable} exceptions: 
in the lexical scope of
$G_i$, we can call \texttt{continue} to resume the part of the computation
after the effect operation (i.e., its continuation).
For instance, the next handler resumes the computation after
handling the first \texttt{out/1} operation and intercepts 
later \texttt{out/1} operations in the same way.
\begin{Verbatim}[framerule=1mm,frame=leftline,xleftmargin=5mm]
  ?- handle hw with (out(X) -> writeln(X), continue).
  hello
  world
  true.
\end{Verbatim}
Interestingly, we can invoke the same continuation multiple times, for
instance both before and after printing the term.
\begin{Verbatim}[framerule=1mm,frame=leftline,xleftmargin=5mm]
  ?- handle hw with (out(X) -> continue, writeln(X), continue).
  world
  hello
  world
  true.
\end{Verbatim}

\paragraph{The \texttt{finally} Clause}
The optional \texttt{finally} clause is performed
when $G_0$ finishes; if omitted, $G_f$ defaults to \texttt{true}. 
\begin{Verbatim}[framerule=1mm,frame=leftline,xleftmargin=5mm]
  ?- handle hw with (out(X) -> writeln(X), continue) 
       finally (writeln(done)).
  hello
  world
  done
  true.
\end{Verbatim}
Note that if the goal does not run to completion, the \texttt{finally} clause
is not invoked.
\begin{Verbatim}[framerule=1mm,frame=leftline,xleftmargin=5mm]
  ?- handle hw with (out(X) -> writeln(X)) 
       finally (writeln(done)).
  hello
  true.
\end{Verbatim}
   
\paragraph{The \texttt{for} Clause}
All variables in the operation and \texttt{finally} clauses are local to that
clause, except if they are declared in the \texttt{for} clause.\footnote{The
\texttt{for} clause plays a similar role as that in Schimpf's \emph{logical
loops}~\cite{logical_loops}.} Every $\mathit{Var}=\mathit{Term}$ pair in the \texttt{for}
clause relates a variable, which we call a \emph{parameter}, that is in scope of all the
operations and \texttt{finally} clauses with a term whose variables are in scope in
the handler context. For instance, the following handler collects
all outputs in a list.
\begin{Verbatim}[framerule=1mm,frame=leftline,xleftmargin=5mm]
  ?- handle hw with (out(X) -> Lin = [X|Lmid], continue(Lmid,Lout))
       finally (Lin=Lout) for (Lin = List, Lout=[]).
  List = [hello,world].
\end{Verbatim}
Note that \texttt{continue} has one argument for each parameter to indicate
which values the parameters take in the continuation.

%

\subsection{Nested Handlers and Forwarding}

Nesting algebraic effect handlers is similar to nesting exception handlers. If
an operation is not ``caught'' by the inner handler, it is \textit{forwarded}
to the outer handler. Moreover, if the inner handler catches an operation and,
in the process of handling it, raises another operation, then this operation is
handled by the outer handler. Let us illustrate both scenarios.

We can easily define a 
non-deterministic choice operator \texttt{or/2} in the style of
\textsc{Tor}~\cite{ppdp2014,DBLP:journals/scp/SchrijversDTD14} in terms of
the primitive \texttt{choice/1} effect which returns either of the two boolean values
\texttt{t} and \texttt{f}. 
\begin{Verbatim}
  :- effect choice/1.

  or(G1,G2) :- choice(B), (B == t -> G1 ; B == f -> G2).

  chooseAny(G) :- handle G with (choice(B) -> (B = t ; B = f), continue). 
\end{Verbatim}
The \texttt{chooseAny} handler interprets \texttt{choice/1} in terms of
Prolog's built-in disjunction \texttt{(;)/2}.
\begin{Verbatim}[framerule=1mm,frame=leftline,xleftmargin=5mm]
  ?- chooseAny(or(X = 1, X = 2)).
  X = 1;
  X = 2.
\end{Verbatim}
To obtain more interesting behavior, we can nest this handler with:
\begin{Verbatim}
  flip(G) :- handle G with (choice(B) -> choice(B1), not(B1,B), continue). 

  not(t,f). not(f,t).
\end{Verbatim}
to flip the branches in a goal without touching the goal's code.
\begin{Verbatim}[framerule=1mm,frame=leftline,xleftmargin=5mm]
  ?- chooseAny(flip(or(X = 1, X = 2))).
  X = 2;
  X = 1.
\end{Verbatim}
What happens is that the inner \texttt{flip} handler intercepts the
\texttt{choice(B)} call of \texttt{or/2}. It produces a new \texttt{choice(B1)}
call that reaches the outer \texttt{chooseAny} handler, and unifies \texttt{B}
with the negation of \texttt{B1}, which affects the choice in the
\texttt{continue}-ation of \texttt{or/2}.

Thanks to forwarding, we can also easily mix different effects. For instance, with:
\begin{Verbatim}
  writeOut(G) :- handle G with (out(T) -> writeln(T), continue). 
\end{Verbatim}
we can combine output and non-determinism.
\begin{Verbatim}[framerule=1mm,frame=leftline,xleftmargin=5mm]
  ?- chooseAny(writeOut(or(out(hello), out(world)))), fail.
  hello
  world
  false.
\end{Verbatim}
Note that the inner \texttt{writeOut} handler does not know how to interpret the 
\texttt{choice/1} effect. As a consequence, it (implicitly) \emph{forwards} this operation to the
next surrounding handler, \texttt{chooseAny}, who does know what to do.

\subsection{Elaboration Semantics}\label{sec:semantics}

There is a straightforward elaboration of handlers into the \texttt{shift/1}
and \texttt{reset/3} delimited control primitives for Prolog~\cite{DBLP:journals/tplp/SchrijversDDW13}. For instance, the last example query of
Section~\ref{sec:syntax} is elaborated into:
\begin{Verbatim}[framerule=1mm,frame=leftline,xleftmargin=5mm]
  ?- handler42(hw,List,[]). 
  List = [hello,world].
\end{Verbatim}
where the declaration of the \texttt{out/1} operation is elaborated into:
\begin{Verbatim}
  out(X) :- shift(out(X)).  
\end{Verbatim}
which shifts the term representation of the operation.  The actual handler code
is elaborated into a predicate (with a \emph{fresh} name).
\begin{Verbatim}
  handler42(Goal,Lin,Lout) :-
    reset(Goal,Cont,Signal),
    ( Signal == 0 ->
        Lin = Lout
    ; Signal = out(X) ->
        Lin = [X|Lmid],
        handler42(Cont,Lmid,Lout)
    ; shift(Signal), 
      handler42(Cont,Lin,Lout)
    ).
\end{Verbatim}
This predicate executes the goal in the delimited scope of a \texttt{reset/3},
which captures any \texttt{shift/1} call. If the goal terminates normally
(i.e., \texttt{Signal=0}), then the \texttt{finally} code is run. If the
goal suspends with a \texttt{shift/1}, the predicate checks whether the operation
matches the handler's operation clause. If so, the clause's body is run. Note that
\texttt{continue(Lmid,Lout)} has been expanded into a recursive invocation of the handler
with the actual continuation goal \texttt{Cont}. If the operation does not match,
the handler forwards it to the nearest surrounding handler with \texttt{shift/1} and
continues with the continuation.

The example above generalizes straightforwardly.
Any declaration of an effect operation 
is elaborated into a predicate definition.

\begin{minipage}{.99\textwidth}
\begin{BVerbatim}[commandchars=\\\{\}]
  :- effect \textit{op}/\textit{n}.
\end{BVerbatim}
\hfill$\mapsto$\hfill
\begin{BVerbatim}[commandchars=\\\{\}]
  \textit{op}(X1,...,X\textit{n}) :- shift(\textit{op}(X1,...,X\textit{n})).
\end{BVerbatim}
\end{minipage}

Also every handler goal is substituted with a predicate call.
\[
\begin{array}{lcr}
          \begin{array}{c}
             \texttt{handle } G_0  \texttt{ with} \\
             op_{1} \rightarrow G_{1};\\
             \dots\\
             op_{n} \rightarrow G_{m}\\
             \texttt{finally($G_f$)} \\
             \texttt{for($P_1$ = $T_1$,\ldots,$P_n$ = $T_n$)}
         \end{array}
         &
         \mapsto
         &
         \texttt{$h$($G_0$,$T_1$,...,$T_n$).}
\end{array}
\]
where $h/n+1$ is an auxiliary predicate defined as:
\begin{Verbatim}[commandchars=\\\{\},codes={\catcode`$=3\catcode`^=7\catcode`_=8}]
  $h$(Goal,$P_1$,..,$P_n$) :-
    reset(Goal,Cont,Signal),
    ( Signal == 0  -> $G_f$
    ; Signal = $\mathit{op}_1$ -> $G_1^\prime$
    ; ...
    ; Signal = $\mathit{op}_n$ -> $G_n^\prime$
    ; shift(Signal), $h$(Cont,$P_1$,...,$P_n$)
    ).
\end{Verbatim}
Here, each $G_i'$ is derived from $G_i$ by
replacing all occurrences of \texttt{continue($S_1$,\ldots,$S_n$)}
with recursive calls \texttt{$h$(Cont,$S_1$,\ldots,$S_n$)}.

\section{Optimisation}

Section~\ref{sec:semantics}'s elaboration of algebraic effects into the
delimited control constructs is conveniently straightforward. Unfortunately,
capturing the delimited continuation incurs a non-trivial runtime cost. In
many simple cases this cost is quite steep compared to more conventional 
program transformation approaches. For instance, the implementation of DCGs
with delimited control is 10 times slower in a tight loop than the traditional 
term expansion approach.

Fortunately, the runtime overhead is not inherent in the algebraic effects and handlers
approach, and we can obtain competitive performance through optimised compilation.
This section presents our optimisation approach, which aims to eliminate most uses
of delimited control.
The optimisation consists of two collaborating transformation approaches:
rewrite rules (Section~\ref{sec:rewriterules}) and partial evaluation (Section~\ref{sec:partialevaluation}). We
use term rewrite rules to simplify handler goals and possibly eliminate the
handler construct altogether.
These rules depend on an effect system (Section~\ref{sec:effectsystem}) that infers which
effects can or cannot be generated by a goal. Partial evaluation complements
the rewrite rules by specialising handled predicate calls. This enables in
particular the specialisation of (mutually) recursive predicates.

\subsection{Effect System}\label{sec:effectsystem}

Driving our optimisation is an \emph{effect system} that associates with each
goal $G$ an \emph{effect set} $E$ that denotes which effects the goal may call.

\paragraph{Effect Sets}
In order to cater for modular programs, effect sets $E$ are not elements of the
powerset lattice over the closed set $\mathit{OP}$ of locally known effect
operation symbols $\mathit{op}/n$.
Instead we use the powerset lattice over an open-ended set of effect operations
augmented with the additional top element $\mathit{All}$. This allows us to
express the effects of unknown goals and unknown effect operations in an
abstract manner.

Hence, we denote effect sets in one of two forms: $\bigcup_i \mathit{op}_i/n_i$
or $\mathit{All} - \bigcup_i \mathit{op}_i/n_i$.  The former is an explicit
enumeration of effect operations, while the latter expresses the dual: all but
the given effect operations.

The $\in$ relation as well as the functions $\cup$ and $-$ are extended
in the obvious way from the powerset lattice to our augmented version.
\paragraph{Effect System}

We use these functions over effect sets in the definition of our effect system
judgment $E_c \vdash G : E$. This judgment expresses that goal $G$ calls only
effect operations from the effect set $E$, provided that \texttt{continue} calls
only effect operations from the effect set $E_c$. Since \texttt{continue} is not
defined for a top-level goal, we may assume any value for its $E_c$. Hence, for convenience,
we always take $E_c = \emptyset$ for top-level goals $G$ and just write $\vdash G : E$.

Figure~\ref{fig:infrules} defines this judgment by means of inference rules.
Rule~\textsc{(E-Var)} expresses that a variable (i.e., unknown) goal may call
$\mathit{All}$ effect operations. Rule~\textsc{(E-Op)} states that known effect operation
calls itself. Rules~\textsc{(E-Conj)} and \textsc{(E-Disj)} combine the effects of their subgoals.
Rule~\textsc{(E-True)} expresses that the goal \texttt{true}, as an example for other built-ins,
is op-free. Rule~\textsc{(E-Cont)} captures the invariant that \texttt{continue} has
the $E_c$ effect. In Rule~\textsc{(E-Pred)} the effect of a user-defined predicate is the effect of its
body. Finally, the most of the complexity of the inference system is concentrated in Rule~\textsc{(E-Handle)}
that deals with a handler goal. The rule expresses that the handler goal
forwards all the effect operations $E_0$ of the goal $G_0$ it handles, except for the ones that
the handler takes care of, $\bigcup_i \mathit{op}_i/n_i$. In addition, the handler may introduce additional
calls to effect operations in its operation and \texttt{finally} clauses. Also note that 
calls to \texttt{continue} in the operation clauses have exactly the same effect as the handler goal
itself; they are essentially recursive calls after all.

Here are a few examples:
\[\begin{array}{l}
\vdash \texttt{hw} : \texttt{out/1} \\
\vdash \texttt{handle hw with (out(X) -> writeln(X))} : \emptyset \\
\vdash \texttt{handle Y with (out(X) -> writeln(X))} : \mathit{All} - \texttt{out/1} \\
\end{array}\]

\begin{figure}[t]
\centering
\[\begin{array}{l@{\hspace{10mm}}l}

       \GAMMAECVD X : \mathit{All}\hspace{2mm}
      \textsc{(E-Var)} 
      &
 
 \prooftree
        \mathit{op}/n \in \mathit{OP}
 \justifies
 	\GAMMAECVD \mathit{op}(T_1,\ldots,T_n) : \mathit{op}/n
    \using
 	\textsc{(E-Op)}
 \endprooftree
 \end{array}
 \]

\[ \begin{array}{c@{\quad\quad}c}
   \inferrule*[right=\textsc{(E-Conj)}]
     {\GAMMAECVD G_1:E_1 \\ \GAMMAECVD G_2:E_2}
     {\GAMMAECVD (G_1,G_2) : E_1 \cup E_2}
   &
   \inferrule*[right=\textsc{(E-Disj)}]
     {\GAMMAECVD G_1:E_1 \\ \GAMMAECVD G_2:E_2}
     {\GAMMAECVD (G_1;G_2) : E_1 \cup E_2}
   \end{array}
 \]

\vspace{2mm}  

\[\begin{array}{c@{\hspace{2cm}}c}
    \GAMMAECVD \texttt{continue}(\bar{T}) : E_c
    \hspace{2mm} 
 	\textsc{(E-Cont)}
   &
 	\GAMMAECVD \texttt{true} : \emptyset
    \hspace{2mm}
 	\textsc{(E-True)}
 
  \end{array}
 \]	
 
 \vspace{2mm} 

 \prooftree
 \begin{array}{lr}
 	p(S_1,\ldots,S_n) \texttt{ :- } G &
 	\GAMMAECVD G:E
 \end{array}
 \justifies
 \GAMMAECVD p(T_1,\ldots,T_n) : E
 \using
  \textsc{(E-Pred)}
 \endprooftree
\vspace{4mm} 
\prooftree
 	\begin{array}{c}
 	E^* =      
	   		(E_0 - \bigcup_i op_i ) \cup E_f \cup \bigcup_i E_i \vspace{1mm}\\
   
	    \begin{array}{lcr}
	 	\GAMMAECVD G_0:E_0   & \GAMMAECVD G_f:E_f & E^* \vdash G_i :E_i \hspace{1mm}
	 	(\forall i)
	 	\end{array}
 	\end{array}
\justifies
   \begin{array}{lcr}
   \GAMMAECVD & 
   
   \left( \begin{array}{c}
  		   \texttt{handle } G_0  \texttt{ with} \\
  		   \overline{op_{i}(\bar{X}) \rightarrow G_{i}}\\
  		   \texttt{finally}(G_f)~\texttt{for}(G_s)
   		  \end{array}
   \right)

   &:E^*
   
    \end{array}
\using
 	\textsc{(E-Handle)}
 \endprooftree

 \vspace{3mm}
\caption{Effect Inference Rules}
\label{fig:infrules}
\end{figure}

\subsection{Rewrite Rules}\label{sec:rewriterules}

We use the information of the effect system and the syntactic structure of
goals to perform a number of handler-specific optimisations. We denote these
optimisations in terms of semantics-preserving equivalences $G_1 \equiv G_2$
that we use as left-to-right rewrite rules. Figure~\ref{fig:opti} lists our
rewrite rules in the form of inference rules where conditions on the inferred
effects are written above the bar.

\begin{figure}[t]
\vspace{1.5mm}

   \[\begin{array}{l@{\hspace{0.1mm}}c@{\hspace{0.1mm}}l}
   \left( \begin{array}{@{}c@{}}
         \texttt{handle ($G_1$;$G_2$)} \arcr \texttt{with }  
         \overline{op \rightarrow G;} \arcr
         \texttt{finally}(G_f)
                   ~\texttt{for} (G_s) 
        \end{array}
   \right)

   & ~~\equiv~~ 
   & \left(  \begin{array}{@{}c@{}}
             \texttt{handle } G_1 \arcr \texttt{with } 
          \overline{op \rightarrow G;} \arcr
         \texttt{finally}(G_f)
                   ~\texttt{for} (G_s) 
       \end{array}
   \right) 
   \texttt{;} 
   \left(  \begin{array}{@{}c@{}}
         \texttt{handle } G_2  \arcr \texttt{with }
        \overline{op \rightarrow G;} \arcr
         \texttt{finally}(G_f)
                   ~\texttt{for} (G_s) 
       \end{array}
   \right) 
   
    \end{array}\]
  \hspace{10cm}\textsc{(O-Disj)}

\vspace{2mm}

\prooftree
  E_c \vdash G_1 : E_1 \quad\quad E_1 \cap  \bigcup_i op_i = \emptyset
\justifies
   \begin{array}{lcr}
   \left( \begin{array}{c}
         \texttt{handle ($G_1$,$G_2$) with} \arcr
         \overline{op \rightarrow G;} \arcr
         \texttt{finally}(G_f)~\texttt{for} (G_s)
        \end{array}
   \right)

   & \equiv
   & G_1 \texttt{,} \left(  \begin{array}{c}
                      \texttt{handle } G_2 \texttt{ with} \arcr
                   \overline{op \rightarrow G;} \arcr
                   \texttt{finally}(G_f)~\texttt{for} (G_s)
                  \end{array}
             \right) 
   
    \end{array}
\using
  \quad\textsc{(O-Conj)}
 \endprooftree

\vspace{4mm}

\prooftree
 (op(\bar{S}) \rightarrow G_i) \in \overline{op \rightarrow G}\quad\quad \mathit{freshen}(\bar{P}_F,\bar{S},G_i) = (\bar{P}_F',\bar{S}',G_i') \hspace{1cm} \textsc{(O-Op)}
\justifies
   \begin{array}{l@{\hspace{0.1mm}}c@{\hspace{0.1mm}}l}
   \left( \begin{array}{c}
         \texttt{handle } ( op(\bar{T}), G_c) \arcr
                    \texttt{with}~\overline{op \rightarrow G} \arcr
         \texttt{finally}(G_f)\arcr
         \texttt{for} (\bar{P}_F \texttt{=} \bar{P}_A)
        \end{array}
   \right)

   & ~\equiv~
   &   \bar{P}_F' \texttt{=} \bar{P}_A\texttt{,} \bar{T} \texttt{=} \bar{S}'\texttt{,} G_i' \left[
         \texttt{continue}(\bar{U}) \mapsto
         \begin{array}{c}
         \texttt{handle } G_c  \texttt{ with} \arcr
                    \texttt{with}~\overline{op \rightarrow G} \arcr
         \texttt{finally}(G_f)\arcr
         \texttt{for} (\bar{P}_F \texttt{=} \bar{U})
        \end{array}
   \right] 
   
    \end{array}
\using
 \endprooftree

\vspace{6mm}

\prooftree
  \begin{array}{lcr}
  E_c \vdash G : E
  & 
  & \overline{\mathit{op}' \rightarrow G'} = (\overline{\mathit{op} \rightarrow G}) \cap E
  \end{array}
\justifies
   \begin{array}{lcr}
   \left( \begin{array}{c}
         \texttt{handle } G  \arcr
       \texttt{with}~\overline{\mathit{op} \rightarrow G}\arcr 
         \texttt{finally}(G_f)~\texttt{for}(G_s)
        \end{array}
   \right)

   & ~~\equiv~~ 
   &    \left( \begin{array}{c}
         \texttt{handle } G  \arcr
       \texttt{with}~\overline{\mathit{op}' \rightarrow G'}\arcr 
         \texttt{finally}(G_f)~\texttt{for}(G_s)
        \end{array}
   \right) 
   
    \end{array}
\using
    \quad\textsc{(O-Drop)}
 \endprooftree

\vspace{2mm}

   \[\begin{array}{l@{\hspace{0.1mm}}c@{\hspace{0.1mm}}l}
   \left( \begin{array}{c}
         \texttt{handle} (G)
         ~\texttt{with} ~\emptyset \arcr
         \texttt{finally}(G_f)~\texttt{for} (G_s)
        \end{array}
   \right)

   & ~~\equiv~~ G\texttt{,} G_s\texttt{,} G_f \hspace{1cm}\textsc{(O-Triv)}
   
    \end{array}\]

\vspace{2mm}

\prooftree
  \begin{array}{c}
        \text{see text} \quad\quad
    \overline{\mathit{op}_2' \rightarrow G_2'} = (\overline{\mathit{op}_2 \rightarrow G_2}) - \overline{\mathit{op}_1}
  \end{array}
\justifies
   \begin{array}{lcr}
   \left( \begin{array}{l}
      \texttt{handle } 
               \left( \begin{array}{c}
           \texttt{handle } G \arcr
         \texttt{with}~\overline{\mathit{op}_1 \rightarrow G_1}\arcr 
           \texttt{finally}(G_{1,f})\arcr\texttt{for}(G_{1,s})
               \end{array}\right)
            \arcr
    \texttt{with}~\overline{\mathit{op}_2 \rightarrow G_2}\arcr 
      \texttt{finally}(G_{2,f}) \arcr \texttt{for}(G_{2,s})
          \end{array}
   \right)

   & \equiv
   &    \left( \begin{array}{c}
         \texttt{handle } G \texttt{ with} \arcr
       \overline{\mathit{op}_1 \rightarrow G_1'}\arcr 
       \overline{\mathit{op}_2' \rightarrow G_2'}\arcr 
         \texttt{finally}~G_{1,f}' \arcr
                   \texttt{for}(G_{1,s}\textbf{,}G_{2,s})
        \end{array}
   \right) 
   
    \end{array}
\using
    \quad\textsc{(O-Merge)}
 \endprooftree

\caption{Optimisation Rules for effect handlers} \label{fig:opti}
\end{figure}
Rule~\textsc{(O-Disj)} captures the fact that effect handling is orthogonal to
disjunction to specialise the branches of a disjunction separately. There
are two rules for conjunction. Rule~\textsc{(O-Conj)} pulls the first goal $G_1$
of a conjunct out of the handler provided that it does not call any of the
handler's operations. This covers both the case were $G_1$ is an op-free goal
and the case where all the operations in $G_1$ are forwarded by the handler.
The second rule for conjunction, Rule~\textsc{(O-Op)}, statically evaluates the special
case where the first goal is an operation dealt with by the handler. This consists
of three parts: 1) the unification of the formal and actual parameters, 2) the unification
of the formal and actual operation arguments, and 3) calling the operation clause's goal.
Note that we substitute all calls to \texttt{continue($\bar{U}$)} (for any
$\bar{U}$) in this last goal with the second conjunct wrapped in the handler; note that the
arguments $\bar{U}$ become the new actual parameters. In the process we are careful to 
\emph{freshen} all the local logical variables that are used.

Rule~\textsc{(O-Drop)} removes spurious operation clauses from the handler; it only retains
those that correspond to operations that the goal may actually call. In the case that no
operation clauses remain, Rule~\textsc{(O-Triv)} dispenses with the handler altogether. This
amounts to unifying the formal and actual parameters and calling the \texttt{finally} goal.

Finally, the most complex rule of all, Rule~\textsc{(O-Merge)}, merges two nested
handlers into one single handler and thereby eliminates expensive forwarding of
operations. At first it might seem trivial to merge two handlers: We simply
merge all the components of the two handlers pairwise. There is an obvious
simplification to perform in the process: we can drop all outer handler's operation clauses that
overlap with any of the inner handler's clauses, as the inner handler
takes precedence over the outer one.

Yet, there is a further subtle issue that have to be taken account in order to
preserve the original semantics. The \texttt{finally} goal $G_{1,f}$ and the
operation clause goals $\overline{\mathit{op}_1}$ may call operations that are
originally intercepted by the outer handler. We have to make sure that this
remains the case. For that reason we adjust those goals to $G_{1,f}'$ and
$\bar{G}_1'$ in the merged handler. Let us explain these adjustments for
the different forms of operation clause goals $G_{1,i}$ that we consider.
\begin{enumerate}
\item The operation goal $G_{1,i}$ is of the form $G_{1,i,a}\texttt{,}\texttt{continue}(\bar{V})$
      where $G_{1,i,a}$ does not contain any call to \texttt{continue}.
      We wrap the initial part of the goal in the outer handler and finally
      proceed with \texttt{continue}.
\[ G_{1,i}' = 
  \left(\begin{array}{c}
  \texttt{handle } G_{1,i,a} \texttt{ with}\arcr
  \overline{\mathit{op}_2 \rightarrow G_2} \arcr
  \texttt{finally}~\texttt{continue}(\bar{V},\bar{P}_{2,F})
  ~\texttt{for}~(\bar{P}_{2,F},\bar{P}_{2,F}')
  \end{array}\right)
\]
\item  The operation goal $G_{1,i}$ does not contain a call to \texttt{continue}.
       In this case we wrap the entire goal in the outer handler
       and make sure to call the outer handler's final goal.
\[ G_{1,i}' = 
  \left(\begin{array}{c}
  \texttt{handle } G_{1,i} \texttt{ with}
  \overline{\mathit{op}_2 \rightarrow G_2} \arcr
  \texttt{finally}~G_{2,f}
  \texttt{ for}~(\bar{P}_{2,F},\bar{P}_{2,F}')
  \end{array}\right)
\]

\end{enumerate}
Similarly, we adapt the final goal $G_{1,f}$ to
\[ G_{1,f}' = 
  \left(\begin{array}{c}
  \texttt{handle } G_{1,f} \texttt{ with }
  \overline{\mathit{op}_2 \rightarrow G_2} \arcr
  \texttt{finally}~G_{2,f} 
  \texttt{ for}~(\bar{P}_{2,F},\bar{P}_{2,F}')
  \end{array}\right)
\]

\subsection{Partial Evaluation}\label{sec:partialevaluation}

We use a custom partial evaluation approach to expose more optimisation
opportunities for the rewrite rules and to deal with recursive predicates.  Our
partial evaluation is targeted at predicate calls that are handled.  Consider
the following simple DCG example that checks if a phrase is a succession of the
terminals \texttt{ab}:
\begin{Verbatim}[xleftmargin=5mm]
  :- effect c/1.
  ab.                         ab :- c(a), c(b), ab. 
  query(Lin) :-
    handle ab with
      (c(X) -> Lin1=[X|Lmid], continue(Lmid,Lout1))
    finally (Lin1 = Lout1) for (Lin1=Lin,Lout1=[]).
\end{Verbatim}
Here we abstract the goal \texttt{handle ab with \ldots} 
into a fresh predicate (say \texttt{ab0/2}), which makes
abstraction of the actual handler parameters. This yields the
new definition of \texttt{query/1}:
\begin{Verbatim}[xleftmargin=5mm]
  query(Lin) :- ab0(Lin,[]).
\end{Verbatim}
At the same time we unfold the definition of \texttt{ab/0} in the newly created
predicate \texttt{ab0/2}. Because \texttt{ab/0} has two clauses, this means that
\texttt{ab0/2} bifurcates similarly.
\begin{Verbatim}
ab0(Lin,Lout) :-                      ab0(Lin,Lout) :- 
 handle true with                      handle (c(a), c(b), ab) with
 (c(X) -> Lin1=[X|Lmid],               (c(X) -> Lin1=[X|Lmid], 
          continue(Lmid,Lout1))                 continue(Lmid,Lout1))
 finally (Lin1 = Lout1)                finally (Lin1 = Lout1)
 for (Lin1=Lin,Lout1=Lout).            for (Lin1=Lin,Lout1=Lout). 
\end{Verbatim}
This unfolding exposes new rewriting opportunities. Using the Rules
\textsc{(O-Drop)} and \textsc{(O-Triv)}, the first clause specialises to 
\texttt{Lin1=Lin, Lout1=Lout, Lin1=Lout1}. In the second clause, a double
use of Rule \textsc{(O-Op)} deals with the \texttt{c/1} operations. This leaves
a recursive invocation of \texttt{ab/0}, wrapped in the handler. Now the partial
evaluation kicks in again, realises that this is a variant of the earlier
specialiation and ties the knot with a recursive call to \texttt{ab0/2}.
After further clean-up of the unifications, we get:
\begin{Verbatim}[xleftmargin=2mm]
ab0(L,L).                   ab0([a,b|Lmid],Lout) :- ab0(Lmid,Lout).
\end{Verbatim}
There is no trace of delimited control left. Moreover, this is precisely the
tight code that the traditional DCG yields.

\section{Evaluation} \label{sec:evaluation}

We evaluate the usefulness of our optimisation approach experimentally on a set
of benchmarks. All results were obtained on an Intel Core i7 with 8 GB RAM
running hProlog 3.2.38 on Ubuntu 14.04. 

The first experiment concerns the \texttt{ab} program of
Section~\ref{sec:partialevaluation}.  Table \ref{table:dcg} lists the timings
(in ms) for different input sizes obtained with three different versions of the
program: the traditional DCG implementation (based on SICStus),  the elaborated
handler implementation and the optimised handler implementation.  Clearly, the
naive use of delimited control slows the program down by more than an order of
magnitude. Fortunately, our optimisation eliminates all uses of delimited
control and matches the traditional implementation's performance.\footnote{
Thanks to more aggressive inlining it is even slightly faster.}

\begin{table}[t]
\footnotesize
 \begin{tabular}{|c|c|c|c|}
 \cline{1-4}
 Input Size &  Traditional & Elaboration & Optimised \\
 \cline{1-4}
$1\times10^3$  &  0 	&  2   & 0 \\
$1\times10^4$  &  1 	&  4   & 1\\ 
$1\times10^5$  &  8 	&  37  & 5 \\
$1\times10^6$  & 32 	&  321 & 29 \\ 
$2\times10^6$  & 67   	&  635 & 58 \\ 
$5\times10^6$  & 150   &  1821 & 146 \\ 
$1\times10^7$  & 300   &  4757 & 297  \\
$1\times10^8$  & 2953   &  47632 & 2922  \\
  \cline{1-4}
 \end{tabular}
 \caption{DCG benchmark results in ms}
 \label{table:dcg}
\end{table}

The second experiment considers three scenarios with nested handlers.
Table \ref{table:benchmarks} lists the runtime results (in ms) for different
input sizes of different versions: the plain elaborated program, the program
optimised with only the rewrite rules and the program optimised with both
rewrite rules and partial evaluation.

The first benchmark, \texttt{state\_dcg}, extends the \texttt{ab} example with
an implicit state that is incremented with every occurrence of \texttt{ab} in
the input. Because the rewrite rules merge the two handlers in this benchmark,
they generate an almost two-fold speed. With partial evaluation, the speed-up
of around two orders of magnitude is much more dramatic. The main reason is
that delimited control is again eliminated.

The second benchmark adds an inner-most dummy handler for an unused \texttt{foo/0}. The aim of
this benchmark is to assess the cost of forwarding. In the plain elaborated version, we can see
there is a significant overhead. Thanks to the rewriting, the three handlers are again merged
and most of the overhead of the spurious handler disappears -- the only remaining cost is the 
spurious \texttt{foo/0} operation clause. Finally, with partial evaluation, all trace of the
\texttt{foo/0} is eliminated.

The third benchmark re-implements the calculator example of Dragan et
al.~\shortcite{structured_state} with two handlers, one to manage an implicit
stack and one to one for an implicit register. The behavior is similar to the
other two benchmarks: merging the handlers roughly halves the runtime and
partially evaluating them speeds up the code by two orders of magnitude.

\begin{table}[ht]
\footnotesize
 \begin{tabular}{|c|c|c|c|c|}
\cline{1-5}
Program Name & Input Size & Elaborated & Rewriting & Rewriting + PE \\
\cline{1-5}

\textit{state\_dcg} &$1\times10^3$  &  3 	&  2  & 0\\ 
\textit{state\_dcg} &$1\times10^4$  &  20 	&  11 & 0\\ 
\textit{state\_dcg} &$1\times10^5$  &  151 	&  63 & 3\\ 
\textit{state\_dcg} &$1\times10^6$  & 1879 	&  604 &  37\\ 
\textit{state\_dcg} &$2\times10^6$  & 2814   &  1208 & 75 \\
\textit{state\_dcg} &$5\times10^6$  & 7919   &  4348  & 186\\ 
\textit{state\_dcg} &$1\times10^7$  & 29695  &  18094 & 375 \\ 
\cline{1-5}
\textit{state\_dcg\_foo} &$1\times10^3$  &  4 	&  3 & 0 \\
\textit{state\_dcg\_foo} &$1\times10^4$  &  23 	&  11 & 0\\ 
\textit{state\_dcg\_foo} &$1\times10^5$  &  358 	&  61 & 3 \\
\textit{state\_dcg\_foo} &$1\times10^6$  & 4666 	&  670 & 37 \\ 
\textit{state\_dcg\_foo} &$2\times10^6$  & 8777   &  1350 & 75 \\ 
\textit{state\_dcg\_foo} &$5\times10^6$  & 30026   &  4551 & 186  \\
\cline{1-5}
\textit{calculator} &$1\times10^3$  &  4 	&  3 &  1\\
\textit{calculator} &$1\times10^4$  &  30 	&  16 & 1\\ 
\textit{calculator} &$1\times10^5$  &  307	&  78 & 10\\ 
\textit{calculator} &$1\times10^6$  & 1195 	&  761& 57 \\
\textit{calculator} &$2\times10^6$  & 3015   &  1525& 110\\ 
\textit{calculator} &$5\times10^6$  & 12326   &  6114& 247 \\
\cline{1-5}
 \end{tabular}
 \caption{Runtimes of nested-handler benchmarks in ms}
 \label{table:benchmarks}
\end{table}
\section{Related Work}

\paragraph{Language Extensions}
Various Prolog language extensions have been proposed in terms of program
transformations. Van Roy has proposed \emph{Extended DCGs}~\cite{edcgs} to
thread multiple named accumulators. Similarly, Ciao Prolog's structured state
threading~\cite{structured_state} enables different implicit states. Algebraic
effects and handlers can easily provide similar functionality. 

Schimpf's \emph{logical loops}~\cite{logical_loops} approach has been very
influential on our handler design, in particular regarding the elaboration into
recursive predicates and the notions of locally fresh variables and parameters.
Of course, both features originate in distinct paradigms: logical loops
are inspired by imperative loops, while handlers originate in the functional
programming paradigm.

\paragraph{Control Primitives}
Various works have considered extensions of Prolog that enable control-flow
manipulation. Before the work of Schrijvers et
al.~\cite{DBLP:journals/tplp/SchrijversDDW13}, Tarau and Dahl already allowed
the users of BinProlog to access and manipulate the program's continuation.

Various coroutine-like features have been proposed in the context of Prolog for
implementing alternative execution mechanisms, such as constraint logic
programming and delay. Nowadays most of these are based on a single primitive
concept: attributed variables
\cite{holzbaurPLILP-92}. 
Like delimited control, attributed variables are a very low-level feature that
is meant to be used directly, but is often used by library writers as the
target for much higher-level declarative features.

\paragraph{Algebraic Effects and Handlers}
The work in this paper adapts the existing work in the functional programming community
on algebraic effects and handlers to Prolog.
Both algebraic effects~\cite{fossacs/PlotkinP02} and
handlers~\cite{plotkinpretnar} were first explored at a theoretical level,
before giving rise to a whole range of implementations in functional
programming languages, such as Eff~\cite{eff}, Multicore OCaml~\cite{ocaml} and
Haskell~\cite{kammar,Kiselyov:2013,WuS15} to name a few.

Schrijvers et al.~\shortcite{ppdp2014} have previously appealed to a functional
model of algebraic effects and handlers to derive a Prolog
implementation of search heuristics~\shortcite{DBLP:journals/scp/SchrijversDTD14}.
This paper enables a direct Prolog implementation that avoids this detour.

\section{Conclusion}

This paper has defined algebraic effects and handlers for Prolog as a
high-level alternative to delimited control for implementing custom
control-flow and dataflow effects.
In order to avoid undue runtime overhead of capturing delimited continuations,
we provide an optimised compilation approach based on partial evaluation and
rewrite rules. Our experimental evaluation shows that this approach greatly
reduces the runtime overhead.

\paragraph{Acknowledgments}
This work is partly funded by the Flemish Fund for Scientific
Research (FWO). We are grateful to Bart Demoen for his support of hProlog.

\bibliographystyle{acmtrans}
\bibliography{biblio}

\begin{thebibliography}{}

\bibitem[\protect\citeauthoryear{Bauer and Pretnar}{Bauer and
  Pretnar}{2015}]{eff}
{\sc Bauer, A.} {\sc and} {\sc Pretnar, M.} 2015.
\newblock Programming with algebraic effects and handlers.
\newblock {\em Journal of Logical and Algebraic Methods in Programming\/}~{\em
  84,\/}~1, 108--123.

\bibitem[\protect\citeauthoryear{Danvy and Filinski}{Danvy and
  Filinski}{1990}]{abstracting_control}
{\sc Danvy, O.} {\sc and} {\sc Filinski, A.} 1990.
\newblock Abstracting control.
\newblock LFP '90. 151--160.

\bibitem[\protect\citeauthoryear{Desouter, van Dooren, and Schrijvers}{Desouter
  et~al\mbox{.}}{2015}]{DBLP:journals/tplp/DesouterDS15}
{\sc Desouter, B.}, {\sc van Dooren, M.}, {\sc and} {\sc Schrijvers, T.} 2015.
\newblock Tabling as a library with delimited control.
\newblock {\em {TPLP}\/}~{\em 15,\/}~4-5, 419--433.

\bibitem[\protect\citeauthoryear{Dijkstra}{Dijkstra}{1968}]{Dijkstra:1968:LEG:362929.362947}
{\sc Dijkstra, E.~W.} 1968.
\newblock Letters to the editor: Go to statement considered harmful.
\newblock {\em Commun. ACM\/}~{\em 11,\/}~3 (Mar.), 147--148.

\bibitem[\protect\citeauthoryear{Dolan, White, Sivaramakrishnan, Yallop, and
  Madhavapeddy}{Dolan et~al\mbox{.}}{2015}]{ocaml}
{\sc Dolan, S.}, {\sc White, L.}, {\sc Sivaramakrishnan, K.}, {\sc Yallop, J.},
  {\sc and} {\sc Madhavapeddy, A.} 2015.
\newblock Effective concurrency through algebraic effects.
\newblock In {\em OCaml Users and Developers Workshop}.

\bibitem[\protect\citeauthoryear{Felleisen}{Felleisen}{1988}]{Felleisen:1988}
{\sc Felleisen, M.} 1988.
\newblock The theory and practice of first-class prompts.
\newblock POPL '88. 180--190.

\bibitem[\protect\citeauthoryear{Holzbaur}{Holzbaur}{1992}]{holzbaurPLILP-92}
{\sc Holzbaur, C.} 1992.
\newblock {Meta-structures vs. Attributed Variables in the Context of
  Extensible Unification}.
\newblock LNCS, vol. 631. 260--268.

\bibitem[\protect\citeauthoryear{Ivanovic, Morales~Caballero, Carro, and
  Hermenegildo}{Ivanovic et~al\mbox{.}}{2009}]{structured_state}
{\sc Ivanovic, D.}, {\sc Morales~Caballero, J.~F.}, {\sc Carro, M.}, {\sc and}
  {\sc Hermenegildo, M.} 2009.
\newblock Towards structured state threading in {P}rolog.
\newblock In {\em CICLOPS 2009}.

\bibitem[\protect\citeauthoryear{Kammar, Lindley, and Oury}{Kammar
  et~al\mbox{.}}{2013}]{kammar}
{\sc Kammar, O.}, {\sc Lindley, S.}, {\sc and} {\sc Oury, N.} 2013.
\newblock Handlers in action.
\newblock In {\em Proceedings of the 18th ACM SIGPLAN International Conference
  on Functional programming}. ICFP '14. ACM, 145--158.

\bibitem[\protect\citeauthoryear{Kiselyov, Sabry, and Swords}{Kiselyov
  et~al\mbox{.}}{2013}]{Kiselyov:2013}
{\sc Kiselyov, O.}, {\sc Sabry, A.}, {\sc and} {\sc Swords, C.} 2013.
\newblock Extensible effects: An alternative to monad transformers.
\newblock In {\em Proceedings of the 2013 ACM SIGPLAN Symposium on Haskell}.
  Haskell '13. ACM, New York, NY, USA, 59--70.

\bibitem[\protect\citeauthoryear{Plotkin and Matija}{Plotkin and
  Matija}{2013}]{plotkinpretnar}
{\sc Plotkin, G.~D.} {\sc and} {\sc Matija, P.} 2013.
\newblock Handling algebraic effects.
\newblock {\em Logical Methods in Computer Science\/}~{\em 9,\/}~4.

\bibitem[\protect\citeauthoryear{Plotkin and Power}{Plotkin and
  Power}{2002}]{fossacs/PlotkinP02}
{\sc Plotkin, G.~D.} {\sc and} {\sc Power, J.} 2002.
\newblock Notions of computation determine monads.
\newblock In {\em Foundations of Software Science and Computation Structures},
  {M.~Nielsen} {and} {U.~Engberg}, Eds. LNCS, vol. 2303. Springer, 342--356.

\bibitem[\protect\citeauthoryear{Roy}{Roy}{1989}]{edcgs}
{\sc Roy, P.~V.} 1989.
\newblock A useful extension to prolog's definite clause grammar notation.
\newblock ~{\em 24,\/}~11, 132--134.

\bibitem[\protect\citeauthoryear{Schimpf}{Schimpf}{2002}]{logical_loops}
{\sc Schimpf, J.} 2002.
\newblock Logical loops.
\newblock In {\em International Conference on Logic Programming}. Springer,
  224--238.

\bibitem[\protect\citeauthoryear{Schrijvers, Demoen, Desouter, and
  Wielemaker}{Schrijvers
  et~al\mbox{.}}{2013}]{DBLP:journals/tplp/SchrijversDDW13}
{\sc Schrijvers, T.}, {\sc Demoen, B.}, {\sc Desouter, B.}, {\sc and} {\sc
  Wielemaker, J.} 2013.
\newblock Delimited continuations for {P}rolog.
\newblock {\em {TPLP}\/}~{\em 13,\/}~4-5, 533--546.

\bibitem[\protect\citeauthoryear{Schrijvers, Demoen, Triska, and
  Desouter}{Schrijvers et~al\mbox{.}}{2014}]{DBLP:journals/scp/SchrijversDTD14}
{\sc Schrijvers, T.}, {\sc Demoen, B.}, {\sc Triska, M.}, {\sc and} {\sc
  Desouter, B.} 2014.
\newblock Tor: Modular search with hookable disjunction.
\newblock {\em Sci. Comput. Program.\/}~{\em 84}, 101--120.

\bibitem[\protect\citeauthoryear{Schrijvers, Wu, Desouter, and
  Demoen}{Schrijvers et~al\mbox{.}}{2014}]{ppdp2014}
{\sc Schrijvers, T.}, {\sc Wu, N.}, {\sc Desouter, B.}, {\sc and} {\sc Demoen,
  B.} 2014.
\newblock Heuristics entwined with handlers combined.
\newblock In {\em Proceedings of the 16th International Symposium on Principles
  and Practice of Declarative Programming}. PPDP '14. ACM, 259--270.

\bibitem[\protect\citeauthoryear{Wu and Schrijvers}{Wu and
  Schrijvers}{2015}]{WuS15}
{\sc Wu, N.} {\sc and} {\sc Schrijvers, T.} 2015.
\newblock Fusion for free - efficient algebraic effect handlers.
\newblock In {\em Mathematics of Program Construction}. LNCS, vol. 9129.
  Springer, 302--322.

\end{thebibliography}

\newpage

\appendix

\section{Detailed Optimisation Example Walk-Through}

This appendix elaborates the optimisation example of Section
\ref{sec:partialevaluation} in more depth.

We start from the following program:
\begin{Verbatim}[xleftmargin=5mm]
  :- effect c/1.
  ab.
  ab :- c(a), c(b), ab. 
  query(Lin) :-
    handle ab with
      (c(X) -> Lin1=[X|Lmid], continue(Lmid,Lout1))
    finally (Lin1 = Lout1)
    for (Lin1=Lin,Lout1=[]).
\end{Verbatim}

\paragraph{Step 1}
We abstract the goal \texttt{handle ab with ...} into
a new predicate \texttt{ab0/2}.  This new predicate takes two arguments: one
for every parameter in the handler's \texttt{for} clause.
The original call is replaced by a call to the new predicate, supplying the actual parameters
of the handler as actual arguments.
\begin{Verbatim}[xleftmargin=5mm]
  query(Lin) :- ab0(Lin,[]).
\end{Verbatim}

The predicate \texttt{ab0/2} is a copy of \texttt{ab/0}'s definition, with the handler wrapped
around each clause's body.
\begin{Verbatim}[xleftmargin=5mm]
  ab0(Lin,Lout) :-
    handle true with
      (c(X) -> Lin1=[X|Lmid], continue(Lmid,Lout1))
    finally (Lin1 = Lout1)
    for (Lin1=Lin,Lout1=Lout).
 
  ab0(Lin,Lout) :- 
    handle (c(a), c(b), ab) with
      (c(X) -> Lin1=[X|Lmid], continue(Lmid,Lout1))
    finally (Lin1 = Lout1)
    for (Lin1=Lin,Lout1=Lout).
\end{Verbatim}

\paragraph{Step 2}
The optimiser now applies rewrite rules to the two clauses. In the first
clause, Rule \textsc{(O-Drop)} can be applied because the effect system
provides the information that the goal \texttt{true} has no effects. Hence,
we drop the operation clause:

\begin{Verbatim}[xleftmargin=5mm]
  ab0(Lin,Lout) :-
    handle true with
    finally (Lin1 = Lout1)
    for (Lin1=Lin,Lout1=Lout).
\end{Verbatim}

\paragraph{Step 3}
The handler currently handles no operations\footnote{This syntax is only allowed during
the compilation process.}. The optimizer proceeds with applying \textsc{(O-Triv)}:

\begin{Verbatim}[xleftmargin=5mm]
  ab0(Lin,Lout) :-
    true, 
    Lin1 = Lout1 ,
    Lin1 = Lin, 
    Lout1 = Lout.
\end{Verbatim}

\paragraph{Step 4}
By partially evaluating \texttt{true} and the remaining unifications, the first clause is
simplified to:
\begin{Verbatim}[xleftmargin=5mm]
  ab0(L,L).
\end{Verbatim} 

\paragraph{Step 5}
In the second clause the handler's goal starts with the \texttt{c/1} operation.
The optimiser applies \textsc{(O-Op)} to the handler, producing the following code:
\begin{Verbatim}[xleftmargin=5mm]
  ab0(Lin,Lout) :- 
    Lin1 = [a|Lmid],
    Lin1 = Lin,
    Lout1 = Lout,
    handle (c(b), ab) with
      (c(X1) -> Lin11=[X1|Lmid1], continue(Lmid1,Lout11))
    finally (Lin11 = Lout11)
    for (Lin11=Lmid,Lout11=Lout1).
\end{Verbatim} 
All the variables in the new handler goal are fresh variables.
Observe that the actual arguments in the newly generated \texttt{for}
clause are taken from the \texttt{continue} call of the previous handler.
This is to ensure the correct state threading of the handler, and to keep the correct
semantics of the program.

\paragraph{Step 6}
The optimiser re-applies \textsc{(O-Op)} for \texttt{c(b)}, generating the following code:
\begin{Verbatim}[xleftmargin=5mm]
  ab0(Lin,Lout) :- 
    Lin1 = [a|Lmid],
    Lin1 = Lin,
    Lout1 = Lout,
    Lin11 = [b|Lmid1],
    Lin11 = Lmid,
    Lout11 = Lout1,
    handle (ab) with
      (c(X2) -> Lin12=[X1|Lmid2], continue(Lmid2,Lout12))
    finally (Lin12 = Lout12)
    for (Lin12=Lmid1,Lout12=Lout11).
\end{Verbatim} 

\paragraph{Step 7}
The remaining handler goal is now a variant of the original one, which was
already abstracted into \texttt{ab0/2}. Therefore, we can replace it
with \texttt{ab0/2}.
\begin{Verbatim} [xleftmargin=5mm]
  ab0(Lin,Lout) :- 
    Lin1 = [a|Lmid],
    Lin1 = Lin,
    Lout1 = Lout,
    Lin11 = [b|Lmid1],
    Lin11 = Lmid,
    Lout11 = Lout1,
    ab0(Lmid1,Lout11).
\end{Verbatim}

\paragraph{Step 8}
The clause now consists of several unifications followed by a tail-recursive call. Partially evaluating the
unifications leads to the final optimised code:
\begin{Verbatim} [xleftmargin=5mm]
  ab0([a,b|Lmid1],Lout) :-
    ab0(Lmid1,Lout).
\end{Verbatim}

\section{State-DCG Handler Example in Detail}

This appendix shows the result of optimizing a program that consists of two handlers. We first show
the elaboration into delimited control. Then, we show how the original program can be optimised
by means of the rewrite rules and partial evaluation.

We use the following program, which was used to generate the results of the first benchmarks in Table 
\ref{table:benchmarks}. As described in Section \ref{sec:evaluation}, there are two handlers:
one handles the implicit state operations and the other handles the DCG operations.

\begin{Verbatim}
abinc.
abinc :- c(a), c(b), get_state(St), St1 is St+1, put_state(St1), abinc.

state_phrase_handler(Sin,Sout,Lin,Lout) :- 
  handle
      (handle abinc 
        with
          ( get_state(Q)  -> Q = Sin1, continue(Sin1,Sout1)
          ; put_state(NS) -> continue(NS,Sout1)
          )
        finally 
          Sout1 = Sin1
        for 
          (Sin1 = Sin, Sout1 = Sout)
      )
    with
      (c(X) -> Lin1 = [X|Lmid], continue(Lmid,Lout1))
    finally 
      Lin1 = Lout1
    for
      (Lin1=Lin, Lout1=Lout).
\end{Verbatim}

The inner handler's goal is \texttt{abinc}, which consumes two elements, \texttt{a}
and \texttt{b}, by using the operation \texttt{c/1} and then increments the state using the 
operations \texttt{get\_state/1} and \texttt{put\_state/1}.

\begin{Verbatim}[framerule=1mm,frame=leftline,xleftmargin=5mm]
  ?- state_phase_handler(0,Sout,[a,b,a,b,a,b],Lout).
  Sout = 0
  Lout = [a,b,a,b,a,b];
  Sout = 1
  Lout = [a,b,a,b];
  Sout = 2
  Lout = [a,b];
  Sout = 3
  Lout = [].
\end{Verbatim}

The immediate elaboration into delimited control yields:
\begin{Verbatim}
state_phrase_handler(A, B, C, D) :-
  handler_0(handler_1(abinc,A,B), C, D).
handler_1(A, B, C) :-
  reset(A, D, E),
  ( D == 0 ->
      C = B
  ; E = get_state(F) ->
      F = B,
      handler_1(D, B, C)
  ; E = put_state(G) ->
      handler_1(D, G, C)
  ;   shift(E),
      handler_1(D, B, C)
  ).
handler_0(A, B, C) :-
  reset(A, D, E),
  ( D == 0 ->
      B = C
  ; E = c(F) ->
      B = [F|G],
      handler_0(D, G, C)
  ;   shift(E),
      handler_0(D, B, C)
  ).
\end{Verbatim}

The predicates \texttt{handler\_0/3} and \texttt{handler\_1/3} correspond to the 
elaborated DCG and state handlers respectively. They follow the semantics described
in Section \ref{sec:semantics}.

Using the rewrite rules first, yields the following elaborated program instead:
\begin{Verbatim}
state_phrase_handler(A, B, C, D) :-
  handler_2(abinc, A, B, C, D).
handler_2(A, B, C, D, E) :-
  reset(A, F, G),
  ( F == 0 ->
      C = B,
      D = E
  ; G = get_state(H) ->
      H = B,
      handler_2(F, B, C, D, E)
  ; G = put_state(I) ->
      handler_2(F, I, C, D, E)
  ; G = c(J) ->
      D = [J|K],
      handler_2(F, B, C, K, E)
  ;   shift(G),
      handler_2(F, B, C, D, E)
  ).
\end{Verbatim}
The two handlers have been merged into one, with the corresponding performance improvement. 

When partial evaluation is enabled as well, the optimisation goes one step further
and yields the following final program:
\begin{Verbatim}
state_phrase_handler(A, B, C, D) :-
  abinc0(A, B, C, D).
abinc0(A, A, B, B).
abinc0(A, B, [a,b|C], D) :-
  E is A+1,
  abinc0(E, B, C, D).
\end{Verbatim}
Partial evluation has pushed the handlers into the definition of \texttt{abcinc/0} where
the rewrite rules have been able to replace the operations by the corresponding handler
clauses. As a consequence, the handlers are eliminated and no delimited control primitives
are generated.

\section{Soundness of Rule \textsc{(O-Disj)}}

This appendix proves the soundness of the \textsc{(O-Disj)} rewrite rule.
Our proof relies on the elaboration of the handler syntax into delimited control
and the corresponding semantics for delimited control given by 
Schrijvers et al.~\shortcite{DBLP:journals/tplp/SchrijversDDW13}. This
semantics is expressed in terms of a Prolog meta-interpreter that we show in
Figure \ref{fig:interpreter}.

\begin{figure}
\begin{Verbatim}[fontsize=\footnotesize,commandchars=\\\{\}]
eval(G) :- eval(G,Signal),
          ( Signal = shift(Term,Cont) ->
            fail
          ; true).          

eval(shift(Term),Signal) :- !,Signal = shift(Term,true).

eval(reset(G,Cont,Term),Signal) :- !, eval(G,Signal1),
                                   ( Signal1 = ok -> Cont = 0, Term = 0
                                   ; Signal1 = shift(Term,Cont)),
                                   Signal = ok.

eval((G1,G2),Signal) :- !, eval(G1,Signal1),
                        ( Signal1 = ok -> eval(G2,Signal)
                        ; Signal1 = shift(Term,Cont),
                          Signal = shift(Term,(Cont,G2))).

eval((G1;G2),Signal) :- !, ( eval(G1,Signal) 
                           ; eval(G2,Signal)).

eval((C->G1;G2),Signal) :- !, ( eval(C,Signal1) ->
                                ( Signal1 = ok -> eval(G1,Signal)
                                ; fail
                                )
                              ; eval(G2,Signal)).

eval(Goal,Signal) :- built_in_predicate(Goal), !, call(Goal), Signal = ok.

eval(Goal,Signal) :- clause(Goal,Body), eval(Body,Signal).
\end{Verbatim}
\caption{Delimited Control Meta-Interpreter} \label{fig:interpreter}
\end{figure}

We start from the left-hand side of the rewrite rule and
turn it into the right-hand side by means of a number of
equivalence preserving transformations.
\begin{equation}
\begin{array}{c}
         \texttt{handle (G1;G2) with} \arcr
         \overline{op \rightarrow G;} \arcr
         \texttt{finally}~G_f \arcr
         \texttt{for}~G_s.
\end{array}
\end{equation}

The elaboration of this handler goal into delimited control yields the following 
auxiliary predicate:

\begin{Verbatim}[commandchars=\\\{\},codes={\catcode`$=3\catcode`^=7\catcode`_=8}]
  $h$(Goal,$P_1$,..,$P_n$) :-
    reset(Goal,Cont,Term),
    ( Term == 0  -> $G_f$
    ; $\mathit{\overline{\texttt{Term = } op \rightarrow G}}$
    ; shift(Signal), $h$(Cont,$P_1$,...,$P_n$)
    ).
\end{Verbatim}

Here the variables $P_i$ are the formal parameters of $G_s$. The goal itself is then
by definition equivalent to 
\begin{equation}
\texttt{h((G1;G2),A$_1$,\ldots,A$_n$)} 
\end{equation}
where the \texttt{A$_i$} are the actual
parameters of $G_s$.

This is equivalent to evaluation the goal in the meta-interpreter:
\begin{equation}
\texttt{eval(h((G1;G2),A$_1$,\ldots,A$_n$))}
\end{equation}

We can now unfold the \texttt{eval/1} call and subsequently unfold the resulting call to the auxiliary predicate \texttt{eval/2}
which selects the last clause. After also evaluating the call to \texttt{clause/2} to unfold \texttt{h/$n+1$} we get:
\begin{equation}
\begin{minipage}{.9\textwidth}
\begin{Verbatim}[commandchars=\\\{\},codes={\catcode`$=3\catcode`^=7\catcode`_=8}]
    eval( ( reset((G1;G2),Cont,Term),
            ( Term == 0  -> $G_f$
            ; $\mathit{\overline{\texttt{Term = } op \rightarrow G}}$
            ; shift(Signal), $h$(Cont,$P_1$,...,$P_n$)
            )
          )
        , Signal
        ),
    (Signal = shift(Term,Cont) -> fail ; true)
\end{Verbatim}
\end{minipage}
\end{equation}
For the sake of space, we refer to the if-then-else block after the
\texttt{reset/3} call as \texttt{<Switches>}. We can thus abbreviate the above as:
\begin{equation}
\begin{minipage}{.9\textwidth}
\begin{Verbatim}[commandchars=\\\{\},codes={\catcode`$=3\catcode`^=7\catcode`_=8}]
    eval( (reset((G1;G2),Cont,Term), <Switches>)
        , Signal
        ),
    (Signal = shift(Term,Cont) -> fail ; true)
\end{Verbatim}
\end{minipage}
\end{equation}
Unfolding \texttt{eval/2} using the appropriate clause for conjunction, yields:
\begin{equation}
\begin{minipage}{.9\textwidth}
\begin{Verbatim}[commandchars=\\\{\},codes={\catcode`$=3\catcode`^=7\catcode`_=8}]
    eval(reset((G1;G2),Cont,Term), Signal1),
    ( Signal1 = ok -> eval(<Switches>, Signal)
    ; Signal1 = shift(Term,Cont) -> Signal = shift(Term,(Cont,<Switches>))
    ),
    (Signal = shift(Term,Cont) -> fail ; true)
\end{Verbatim}
\end{minipage}
\end{equation}
Now we unfold the first call to \texttt{eval/2} using the clause for \texttt{reset/3}:
\begin{equation}
\begin{minipage}{.9\textwidth}
\begin{Verbatim}[commandchars=\\\{\},codes={\catcode`$=3\catcode`^=7\catcode`_=8}]
    eval((G1;G2), Signal2),
    ( Signal2 = ok -> Cont = 0, Term = 0
    ; Signal2 = shift(Term,Cont)
    ),
    Signal1 = ok,
    ( Signal1 = ok -> eval(<Switches>, Signal)
    ; Signal1 = shift(Term,Cont) -> Signal = shift(Term,(Cont,<Switches>))
    ),
    (Signal = shift(Term,Cont) -> fail ; true)
\end{Verbatim}
\end{minipage}
\end{equation}
Again, we unfold the first call to \texttt{eval/2} using the clause for disjunction:
\begin{equation}
\begin{minipage}{.9\textwidth}
\begin{Verbatim}[commandchars=\\\{\},codes={\catcode`$=3\catcode`^=7\catcode`_=8}]
    ( eval(G1, Signal2) ; eval(G2, Signal2) ),
    ( Signal2 = ok -> Cont = 0, Term = 0
    ; Signal2 = shift(Term,Cont)
    ),
    Signal1 = ok,
    ( Signal1 = ok -> eval(<Switches>, Signal)
    ; Signal1 = shift(Term,Cont) -> Signal = shift(Term,(Cont,<Switches>))
    ),
    (Signal = shift(Term,Cont) -> fail ; true)
\end{Verbatim}
\end{minipage}
\end{equation}

We now distribute what comes after the first disjunction into both branches.
\begin{equation}
\begin{minipage}{.9\textwidth}
\begin{Verbatim}[commandchars=\\\{\},codes={\catcode`$=3\catcode`^=7\catcode`_=8}]
    ( 
      eval(G1, Signal2),
      ( Signal2 = ok -> Cont = 0, Term = 0
      ; Signal2 = shift(Term,Cont)
      ),
      Signal1 = ok,
      ( Signal1 = ok -> eval(<Switches>, Signal)
      ; Signal1 = shift(Term,Cont) -> Signal = shift(Term,(Cont,<Switches>))
      ),
      (Signal = shift(Term,Cont) -> fail ; true)
    ; 
      eval(G2, Signal2),
      ( Signal2 = ok -> Cont = 0, Term = 0
      ; Signal2 = shift(Term,Cont)
      ),
      Signal1 = ok,
      ( Signal1 = ok -> eval(<Switches>, Signal)
      ; Signal1 = shift(Term,Cont) -> Signal = shift(Term,(Cont,<Switches>))
      ),
      (Signal = shift(Term,Cont) -> fail ; true)
    )
\end{Verbatim}
\end{minipage}
\end{equation}

At this point we change gear and start folding again. First we fold the
\texttt{reset/2} clause of \texttt{eval/2} twice, once in each branch.
\begin{equation}
\begin{minipage}{.9\textwidth}
\begin{Verbatim}[commandchars=\\\{\},codes={\catcode`$=3\catcode`^=7\catcode`_=8}]
    ( 
      eval(reset(G1,Term,Cont),Signal1),
      ( Signal1 = ok -> eval(<Switches>, Signal)
      ; Signal1 = shift(Term,Cont) -> Signal = shift(Term,(Cont,<Switches>))
      ),
      (Signal = shift(Term,Cont) -> fail ; true)
    ; 
      eval(reset(G2,Term,Cont),Signal1),
      ( Signal1 = ok -> eval(<Switches>, Signal)
      ; Signal1 = shift(Term,Cont) -> Signal = shift(Term,(Cont,<Switches>))
      ),
      (Signal = shift(Term,Cont) -> fail ; true)
    )
\end{Verbatim}
\end{minipage}
\end{equation}

Then we fold the
conjunction clause of \texttt{eval/2} in each branch.
\begin{equation}
\begin{minipage}{.9\textwidth}
\begin{Verbatim}[commandchars=\\\{\},codes={\catcode`$=3\catcode`^=7\catcode`_=8}]
    ( 
      eval((reset(G1,Term,Cont),<Switches>),Signal),
      (Signal = shift(Term,Cont) -> fail ; true)
    ; 
      eval((reset(G2,Term,Cont),<Switches>),Signal),
      (Signal = shift(Term,Cont) -> fail ; true)
    )
\end{Verbatim}
\end{minipage}
\end{equation}

Subsequently, we fold \texttt{eval/1} twice.
\begin{equation}
\begin{minipage}{.9\textwidth}
\begin{Verbatim}[commandchars=\\\{\},codes={\catcode`$=3\catcode`^=7\catcode`_=8}]
    ( 
      eval((reset(G1,Term,Cont),<Switches>))
    ; 
      eval((reset(G2,Term,Cont),<Switches>))
    )
\end{Verbatim}
\end{minipage}
\end{equation}

Now we can drop the meta-interpretation layer again.
\begin{equation}
\begin{minipage}{.9\textwidth}
\begin{Verbatim}[commandchars=\\\{\},codes={\catcode`$=3\catcode`^=7\catcode`_=8}]
    ( 
      (reset(G1,Term,Cont),<Switches>)
    ; 
      (reset(G2,Term,Cont),<Switches>)
    )
\end{Verbatim}
\end{minipage}
\end{equation}

Then we fold \texttt{h/$n+1$} twice.
\begin{equation}
\texttt{( h(G1,A$_1$,\ldots,A$_n$); h(G2,A$_1$,\ldots,A$_n$))} 
\end{equation}

Finally, we invert the elaboration to obtain the right-hand side of the rewrite rule.
\begin{equation}
    \begin{array}{lcr}
        \begin{array}{c}
             \texttt{handle } G_1  \texttt{ with} \arcr
          \overline{op \rightarrow G;} \arcr
         \texttt{finally}(G_f) \arcr
                   \texttt{for} (G_s) 
       \end{array}  &
   \texttt{;} &
    \begin{array}{c}
         \texttt{handle } G_2  \texttt{ with} \arcr
        \overline{op \rightarrow G;} \arcr
         \texttt{finally}(G_f) \arcr
                   \texttt{for} (G_s) 
       \end{array}
  \end{array} 
\end{equation}




\end{document}